# The Peacock Encryption Method

Author: Antti Alexander Kestilä, Helsinki University of Technology, akestila@cc.hut.fi

***Abstract***

*Here is described a preliminary method that enables secure "anti-search-engine" encryption, where the middleman can participate in the encrypted information exchange, without being able to understand the exchanged information, encrypted using a one-way function, as well as being unaware of one of two main exchange participants.*

The Peacock Encryption (PE) method starts from the assumption of a possibility of encryption through a one-way function. This assumption is critical, and implies the use of cryptographic (collision-free) hash functions, such as hash – based identity verification [Lubbe, 1998], when a user-inserted plaintext password is hashed, and compared with the stored hashed password. As seen in Figure 1, Adam and Eve both use the same keyword, which they encrypt with the same algorithm and one-way key, taken from a trusted third party (TTP). Adam then tasks the middleman to search for this same encrypted keyword, and as now the middleman is unable to decipher the meaning of the encryption, he will be able to find Eve's encrypted keyword, but not the keyword itself, and has points Adam to it.

Another important aspect of this method is the masking of one of the two exchange participants. This is essential, because if the middleman is aware of Adam seeking Eve, he has a clear route for a keyword oracle. In order to prevent this, a second enciphered message will be attached to the first one-way keyword encipher (again see Figure 1). The second enciphered message contains pointing address to Eve (Eve's name), and is encrypted in with an independent, reversible encryption algorithm. The deciphering key to this second encipher is then the initial keyword, which both Adam and Eve one-way enciphered, in order to disguise it from the searching middleman. As Adam is pointed to this "feather", he will be able to acquire the second encrypted package, decipher it, and read the correct pointing address. In addition to adding a pointing address to the peacock feather, the two encrypted pieces of information should not be in the possession of Eve in the first place, which would trivialize the encryption of the pointing address. Instead, Eve (and anyone else) should place the feather in a special search site, so as to avoid direct detection by the middleman. Indeed, this search site need (and should) not have anything else in it, besides the keyword feathers.

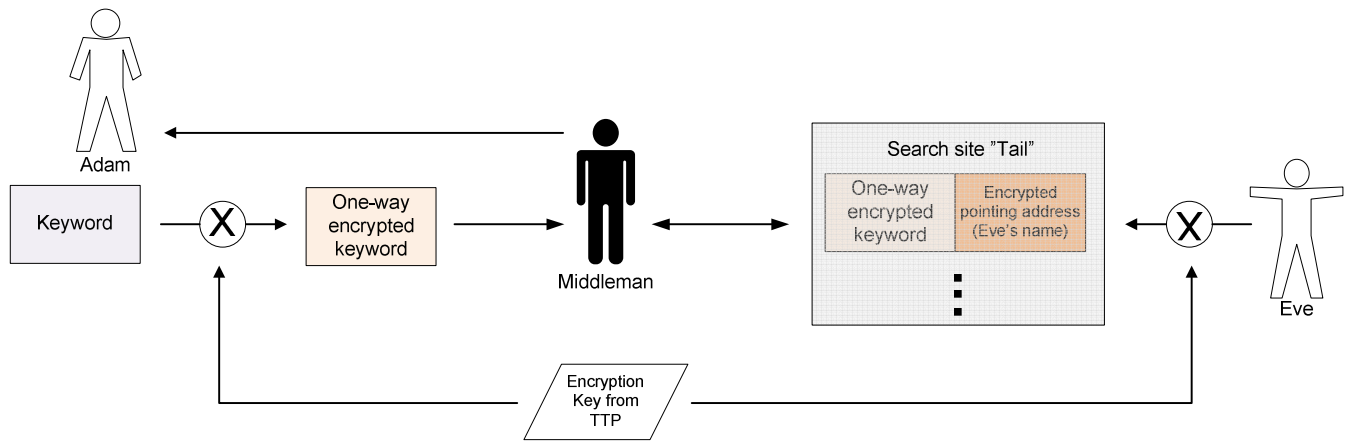

**Figure 1:** **A visual depiction of the exchange protocol.**

This way the PE-method is possible, but it has a considerable weakness, which require more elaboration. Obviously, even if secure encryption through a one-way function were possible, which is widely believed not to be the case [Wenbo, 2003], the middleman himself could use this same method to feed in enough keywords, as to be able to 'map' the encrypted feathers and their pointing addresses. For this reason, the TTP should change the encryption key often enough so as to make mapping or deciphering attempts impractical. This would involve regular checking with the TTP by Adam and Eve, but the information flow would be strictly one-way only – i.e. the TTP will simply announce it publicly, and method users would subsequently copy it down.

This way the middleman, with potentially large computational resources, is included in the search-process without having a possibility of knowing who is searching, and for what.

# Bibliography

C.A. Van Der Lubbe, J. (1998). *Basic Methods of Cryptography.* University of Cambridge Press Syndicate.

Wenbo, M. (2003). *Modern Cryptography: Theory and Practice.* Prentice Hall PTR.